%% file: ms.tex
 \documentclass[12pt,preprint]{aastex}

\slugcomment{in preparation for ApJ}

\shorttitle{WR\,110: A Single Rotating WR Star With CIRs?}
\shortauthors{Chen\'e et al.}

\begin{document}


\title{WR\,110: A Single Wolf-Rayet Star With Corotating Interaction Regions In Its Wind?\footnote{Based on data from the {\it MOST} satellite, a Canadian Space Agency mission, jointly operated by Dynacon Inc., the University of Toronto Institute for Aerospace Studies and the University of British Columbia, with the assistance of the University of Vienna.}}

\author{A.-N. CHEN\'E\altaffilmark{*,\dagger,\ddagger}}
\affil{Canadian Gemini Office, Herzberg Institute of Astrophysics, \\
5071, West Saanich Road, Victoria (BC), V9E 2E7, Canada}
\affil{Departamento de Astronom\'ia, Casilla 160, Universidad de Concepci\'on, Chile}
\affil{Departamento de F\'isica y Astronom\'ia, Facultad de Ciencias,\\ 
Universidad de Valpara\'iso, Av. Gran Breta\~na 1111, Playa Ancha,\\
Casilla 5030, Valpara\'iso, Chile}
\email{achene@astro-udec.cl}

\author{A.~F.~J. MOFFAT} 
\affil{D\'epartement de physique, Universit\'e de Montr\'eal, \\
C.P. 6128, Succ. Centre-Ville, Montr\'eal, QC, H3C 3J7 \& \\
Centre de Recherche en Astrophysique du Qu\'ebec, Canada}
\email{moffat@astro.umontreal.ca}

\author{C. CAMERON} 
\affil{Dept. of Physics \& Astronomy, Univ. of British Columbia, \\
6224 Agricultural Rd., Vancouver, BC, V6T 1Z1, Canada}

\author{R. FAHED\altaffilmark{\ddagger,\amalg}} 
\affil{D\'epartement de physique, Universit\'e de Montr\'eal, \\
C.P. 6128, Succ. Centre-Ville, Montr\'eal, QC, H3C 3J7 \& \\
Centre de Recherche en Astrophysique du Qu\'ebec, Canada}

\author{R.~C. GAMEN\altaffilmark{\amalg}} 
\affil{Instituto de Astrof\'isica de La Plata (CCT-CONICET, \\
Universidad Nacional de La Plata), Paseo del Bosque s/n, 1900, La Plata, Argentina}

\author{L. LEF\`EVRE} 
\affil{Observatoire Royal de Belgique, \\
Avenue Circulaire, 3, 1180 Brussels, Belgium}

\author{J.~F. ROWE}
\affil{NASA Ames Research Center, Moffett Field, CA 94035}

\author{N. ST-LOUIS\altaffilmark{\dagger}} 
\affil{D\'epartement de physique, Universit\'e de Montr\'eal, \\
C.P. 6128, Succ. Centre-Ville, Montr\'eal, QC, H3C 3J7 \& \\
Centre de Recherche en Astrophysique du Qu\'ebec, Canada}

\author{V. MUNTEAN\altaffilmark{\ddagger}} 
\affil{D\'epartement de physique, Universit\'e de Montr\'eal, \\
C.P. 6128, Succ. Centre-Ville, Montr\'eal, QC, H3C 3J7 \& \\
Centre de Recherche en Astrophysique du Qu\'ebec, Canada}

\author{A. DE LA CHEVROTI\`ERE\altaffilmark{\ddagger}} 
\affil{D\'epartement de physique, Universit\'e de Montr\'eal, \\
C.P. 6128, Succ. Centre-Ville, Montr\'eal, QC, H3C 3J7 \& \\
Centre de Recherche en Astrophysique du Qu\'ebec, Canada}

\author{D.~B. GUENTHER} 
\affil{Dept. of Astronomy \& Physics, Saint Mary's University, \\
Halifax, NS, B3H 3C3, Canada}

\author{R. KUSCHNIG} 
\affil{Institut f\"ur Astronomie, Universit\"at Wien, \\
T\"urkenschanzstrasse 17, A-1180 Vienna, Austria}

\author{J.~M. MATTHEWS} 
\affil{Dept. of Physics \& Astronomy, Univ. of British Columbia, \\
6224 Agricultural Rd., Vancouver, BC, V6T 1Z1, Canada}

\author{S.~M. RUCINSKI} 
\affil{Dept. of Astronomy \& Astrophysics, Univ. of Toronto, \\
Toronto, ON, M5S 3H4, Canada}

\author{D. SASSELOV} 
\affil{Harvard-Smithsonian Center for Astrophsics,\\
60 Garden St., Cambridge, MA 102138, USA}

\and

\author{W.~W. WEISS} 
\affil{Institut f\"ur Astronomie, Universit\"at Wien, \\
T\"urkenschanzstrasse 17, A-1180 Vienna, Austria}

\altaffiltext{*}{Visiting Astronomer, Cerro Tololo Inter-American
Observatory. CTIO is operated by AURA, Inc.\ under contract to the
National Science Foundation.}
\altaffiltext{$\dagger$}{Visiting Astronomer at the Canada-France-Hawaii
Telescope (CFHT) which is operated by the National Research Council of
Canada, the Institut National des Sciences de l'Univers of the Centre
National de Recherche Scientifique of France, and the University of
Hawaii.}
\altaffiltext{$\ddagger$}{Visiting Astronomer at the Observatoire du Mont
M\'egantic which is operated by the Centre de Recherche en
Astrophysique du Qu\'ebec.}
\altaffiltext{$\amalg$}{Visiting Astronomer at CASLEO which
is operated under agreement between CONICET and the National Universities of La Plata, C\'ordoba and San Juan, Argentina.}

\begin{abstract}
  A 30-day contiguous photometric run with the {\it MOST} satellite on the WN5-6b star WR\,110 (HD 165688) reveals a fundamental periodicity of P = 4.08 $\pm$ 0.55~days along with a number of harmonics at periods P/n, with n $\approx$ 2,3,4,5 and 6, and a few other possible stray periodicities and/or stochastic variability on timescales longer than about a day. Spectroscopic RV studies fail to reveal any plausible companion with a period in this range. Therefore, we conjecture that the observed light-curve cusps of amplitude $\sim$ 0.01 mag that recur at a 4.08 day timescale may arise in the inner parts, or at the base of, a corotating interaction region (CIR) seen in emission as it rotates around with the star at constant angular velocity. The hard X-ray component seen in WR\,110 could then be a result of a high velocity component of the CIR shock interacting with the ambient wind at several stellar radii. Given that most hot, luminous stars showing CIRs have two CIR arms, it is possible that either the fundamental period is 8.2d or, more likely in the case of WR\,110, there is indeed a second weaker CIR arm for P = 4.08d, that occurs $\sim$two thirds of a rotation period after the main CIR. If this interpretation is correct, WR\,110 therefore joins the ranks with three other single WR stars, all WN, with confirmed CIR rotation periods (WR\,1, WR\,6, and WR\,134), albeit with WR\,110 having by far the lowest amplitude photometric modulation. This illustrates the power of being able to secure intense, continuous high-precision photometry from space-based platforms such as {\it MOST}. It also opens the door to revealing low-amplitude photometric variations in other WN stars, where previous attempts have failed.  If all WN stars have CIRs at some level, this could be important for revealing sources of magnetism or pulsation in addition to rotation periods.
\end{abstract}


\section{Introduction}

While most adequately observed O stars exhibit pairs of opposing Corotating Interaction Regions \citep[CIRs; see ][]{Mu84}, manifested by propagating Discrete Absorption Components in their resonant P Cygni absorption components \citep{Ka99}, little such evidence is found for CIRs in O-star descendants, the Wolf-Rayet (WR) stars. If found in WR stars, CIRs would give precious information on their rotation, which would be highly relevant in the context of rapidly rotating WR stars as prime candidates for slow Gamma-Ray Bursts \citep{St09}.

Recently \citet{St09} (see also \citet{Ch10}) have examined a northern sample of 25 WR stars for the presence of CIRs via large-scale variations exhibited on broad, strong emission lines. Some 20\% of these stars show clear signatures of such effects, with WR\,110 situated at the limit between those stars that show CIRs and those that do not. In addition, some (and possibly all, given the limited number of spectra) stars in the sample unsurprisingly show small-scale wind-clump variations. Such variations manifest themselves as stochastic propagating small-scale spectral variations on time series of virtually all WR line spectra \citep[e.g.][]{Le99}.

WR\,110 is a moderately bright ($v = 10.30, b-v = 0.75$), reddened ($A_v = 3.83$ mag), southern [RA(2000) = 18:07:56.96, DEC(2000) = -19:23:56.8] Galactic ($l = 10.80^\circ,\, b = +0.39^\circ$) Wolf-Rayet (WR) star in the mid-nitrogen sequence with type WN5-6b \citep{Sm96}. The ``b'' suffix indicates broad lines for its subtype and thus low H-abundance by default. Located at a distance of 1.28~kpc, it is~9 pc above the Galactic plane \citep{va01}, in the normal range for most extreme population I stars. Although to date it lacks any obvious binary signature, a comprehensive radial velocity (RV) search has not yet been carried out. However, WR\,110 does have relatively high X-ray flux, with a significant hard component above 3 keV compared to its lower-energy 0.5 keV emission, as seen in most WN stars \citep{Sk02}. The lower-energy component likely arises due to small-scale turbulent shocks in the wind (assuming thermalization of matter colliding at velocity $v$: $kT \approx 1/2 m_p v^2$), with velocity dispersion $\sim\, 10^2$~km~s$^{-1}$, as seen in virtually all WR-star winds \citep{Le99} and other luminous hot stars \citep[e.g.][]{Ev98,Le08}. The former hard component implies velocities of $\sim$10$^3$~km~s$^{-1}$, which are more difficult to explain, without invoking an additional mechanism to provide such high speeds (e.g. accretion onto a low-mass companion or colliding winds with a massive companion). However, WR\,110 emits normal thermal radio emission from its wind \citep{Sk02}, which lends little support for its binary nature. As we show in this investigation, it is possible that  CIRs are present in the wind of WR\,110, and that WR\,110's hard X-rays may arise in the shocks produced by its CIRs at velocities of $\sim$10$^3$~km~s$^{-1}$. 

\section{Observations}

The optical observations of this work consist of both high-precision, nearly continuous photometry over a month ($T_{ph} =$ 30 days) from space using the Microvariability and Oscillations of STars ({\it MOST}) microsatellite, and ground-based optical spectroscopy from various observatories over 8~years.

\subsection{Photometry}

{\it MOST} \citep{Ma99,Wa03} contains a 15-cm Rumak-Maksutov telescope imaging onto a CCD detector via a custom optical broadband (350-750~nm) filter. From its polar Sun-synchronous orbit of altitude 820~km and period 101 min, {\it MOST} has a continuous viewing zone (CVZ) about 54$^\circ$ wide within which it can monitor target fields for up to two months without interruption. Targets brighter than V $\sim$ 6 mag are observed in Fabry imaging mode, while fainter targets (like WR\,110) are observed in direct imaging mode (3$"$ x 3$"$ pixels in a $\sim 1^\circ$ field), similar to standard CCD photometry with a ground-based instrument. The photometry is non-differential, but given the orbit, thermal and design characteristics of {\it MOST}, experience has shown that it is a very photometrically stable platform even over long timescales (with repeatability of the mean instrumental flux from a non-variable target of the brightness of WR\,110 to within about 1~mmag over a month).

Due to the failure of the tracking CCD, the science exposures on the {\it MOST} science CCD now take place at the same time as the guide-star exposures for satellite pointing. In the case of the WR\,110 observations, the guide-star exposure time (and hence also the science target exposure time) was 0.3~sec. To build up sufficient signal-to-noise ratio (S/N), the science exposures were co-added on board the satellite in ``stacks'' of 100 exposures. Stacks (each 30~sec of total integration) were downloaded from the satellite consecutively (with no dead time between stacks), giving a sampling rate of about twice per minute.

The WR\,110 field lies outside the {\it MOST} science CCD CVZ, so it could not be monitored (strictly speaking) continuously. This field alternated with another {\it MOST} primary science target (Altair) during each {\it MOST} satellite orbit. Typically, we monitored WR\,110 for about 40-55~\% of each {\it MOST} orbit (i.e. about 40-55 min of each 101 min) during the low-straylight portion of the orbit, with $\sim$~35 000 individual 30-second data-points between 2007 June 17 and July 16 (HJD 2~454~267 -- 297).

Figure~\ref{fig1} shows the entire {\it MOST} light curve, binned for each {\it MOST} orbit. Although significant variations are seen on timescales longer than $\sim$a day, few variations are seen on shorter timescales and binning allows one to increase the photometric point-to-point {\it rms} precision from $\sim$3 mmag before binning to $\sim$0.5 mmag after binning. Interpretation of the observed variations is given below in the next section.

\subsection{Spectroscopy}

Nine CCD-spectroscopic runs were carried out at five telescopes over eight years, including one parallel run with the {\it MOST} observations. However, this run was plagued by bad weather and instrumental problems, resulting in limited time coverage. A journal of observations is given in Table~\ref{tab1}. Generally, the spectra cover timescales ranging from $\sim$10 minutes to several years and range from near the Balmer limit to He{\sc i}$\lambda$5876, although not the same for each telescope. A SNR$\sim$100 was obtained per extracted pixel, which ranged from 0.5 to 1.6 \AA, with resolution of 2-3 pixels. The bias subtraction, flat-fielding, spectrum extraction, sky subtraction and wavelength calibration of all spectra were executed in the usual way using {\sc iraf}\footnote{{\sc iraf} is distributed by the National Optical Astronomy Observatories (NOAO), which is operated by the Association of Universities for Research in Astronomy, Inc. (AURA) under cooperative agreement with the National Science Foundation (NSF).} software.

Special care was taken for the normalization of the spectra. First, a mean was made for each run. Then each spectrum of a run was divided by the run mean and the ratio fitted with a low order Legendre polynomial (between 4$^{th}$ and 8$^{th}$ order). The original individual spectrum was divided by this fit, and was therefore normalized to the run mean. When this procedure was done for each run, the run means were then put at the same level by using the same procedure as described above. This allowed us to put all individual spectra at the same level. Then, we combined all run means into a global, high quality mean, which was then fitted in selected pseudo-continuum regions, i.e. wavelength regions where large emission-lines do not dominate. These regions are shown in Figure~\ref{fig2}. Finally, the fitted continuum function is applied to rectify each individual spectrum. The error on the continuum normalization -- measured as the standard deviation of individual spectra around the continuum function -- is typically of 0.5\%. Mean spectra from each telescope are shown in Figure~\ref{fig2}.

\section{Analysis}

\subsection{Light Curve}

\begin{figure}[ht]
 \plotone{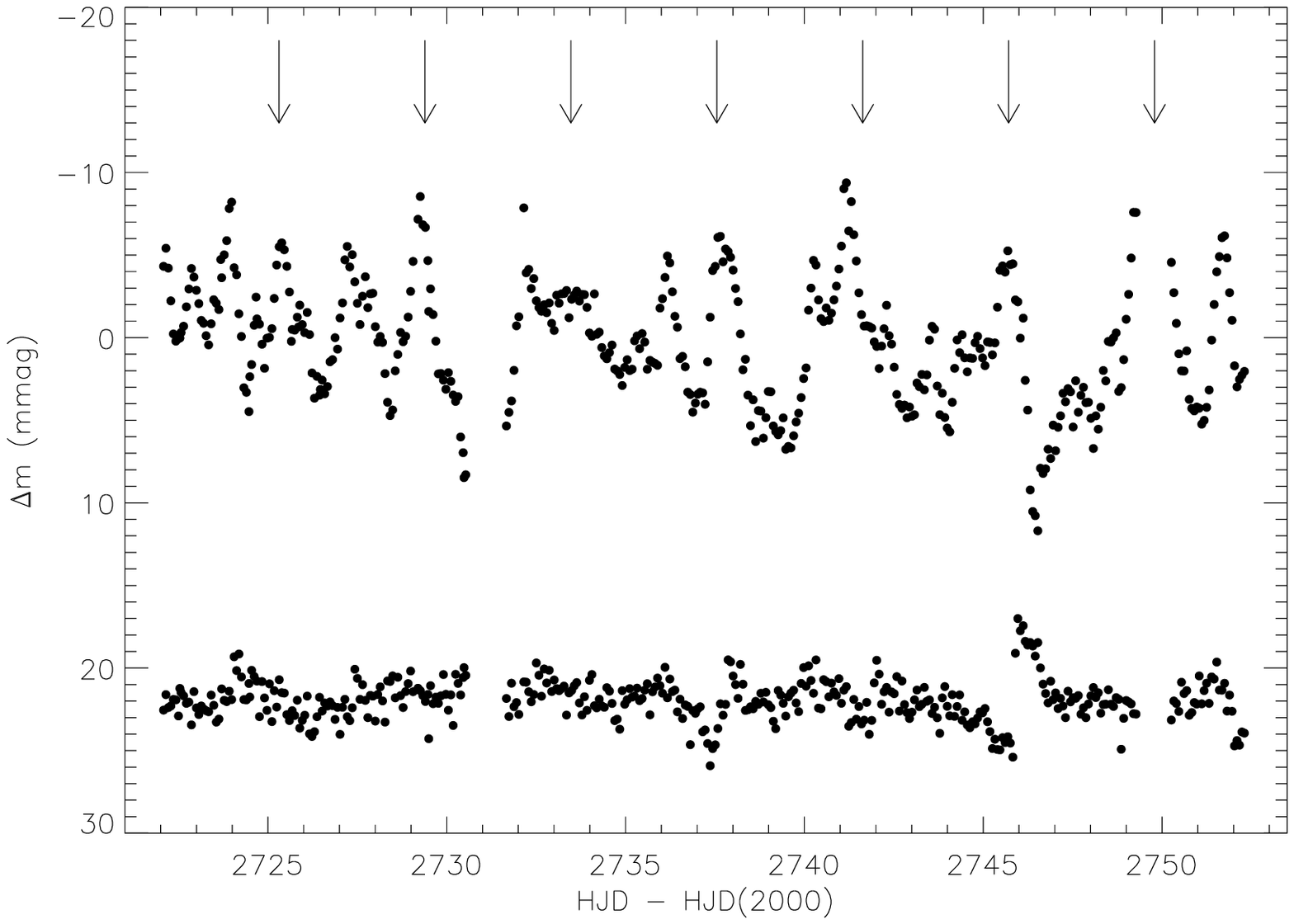}
 \caption{{\it Upper}: {\it MOST} light curve of WR\,110 in 101-minute ({\it MOST} orbit) bins. The vertical lines indicate intervals of period 4.08 d starting at $t_0 =$ HJD~2454270.27. {\it Lower}: Simultaneous {\it MOST} light curve of HD\,312657,  V = 10.8, Sp = A2, reduced in the same way as WR\,110, with arbitrarily displaced magnitude zero-point. This star appears to be stochastically variable at the 0.005 mag peak-to-valley level and slightly larger point-to-point rms scatter of $\sim$1 mmag, compared to $\sim$0.5 mmag for WR\,110.}\label{fig1}
\end{figure}

\begin{figure}[ht]
 \plotone{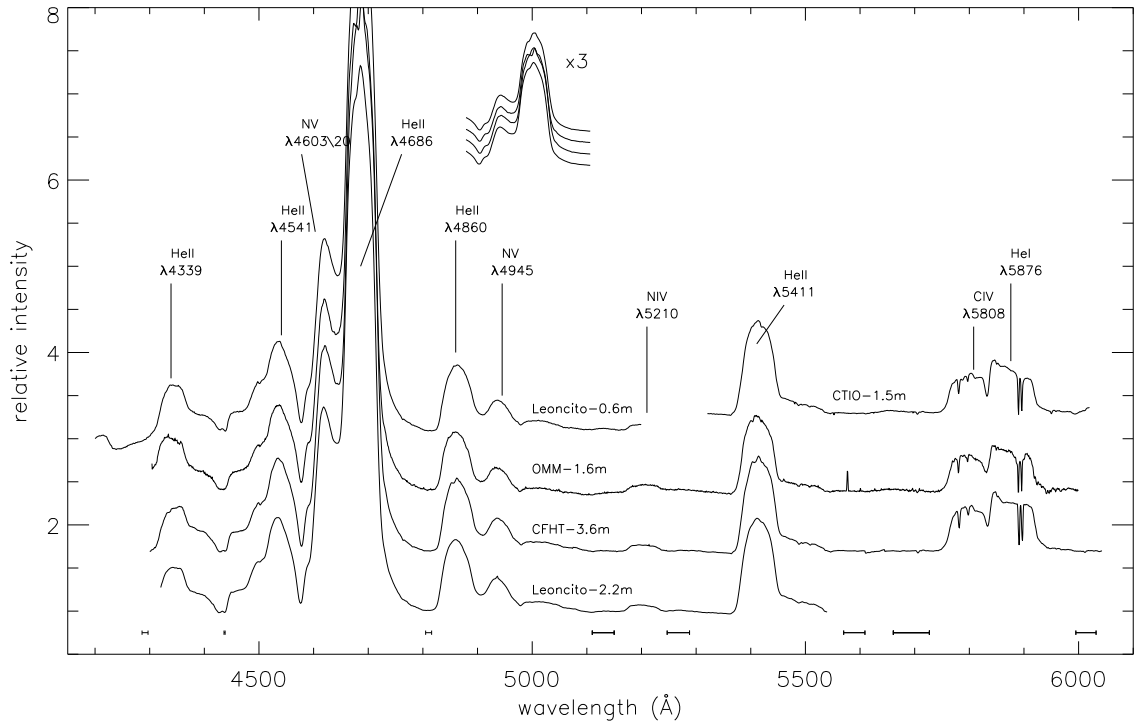}
 \caption{Mean spectra from each run as in Table~\ref{tab1}, with some key lines identified. The vertical scale refers to the lowest spectrum, with the remaining spectra vertically displaced for clarity. The insert shows a vertical compression of the strong He{\sc ii}$\lambda$4686-line region. The horizontal bars show positions used to calculate the continuum.} \label{fig2}
\end{figure}

\begin{figure}[ht]
 \epsscale{.8}
 \plotone{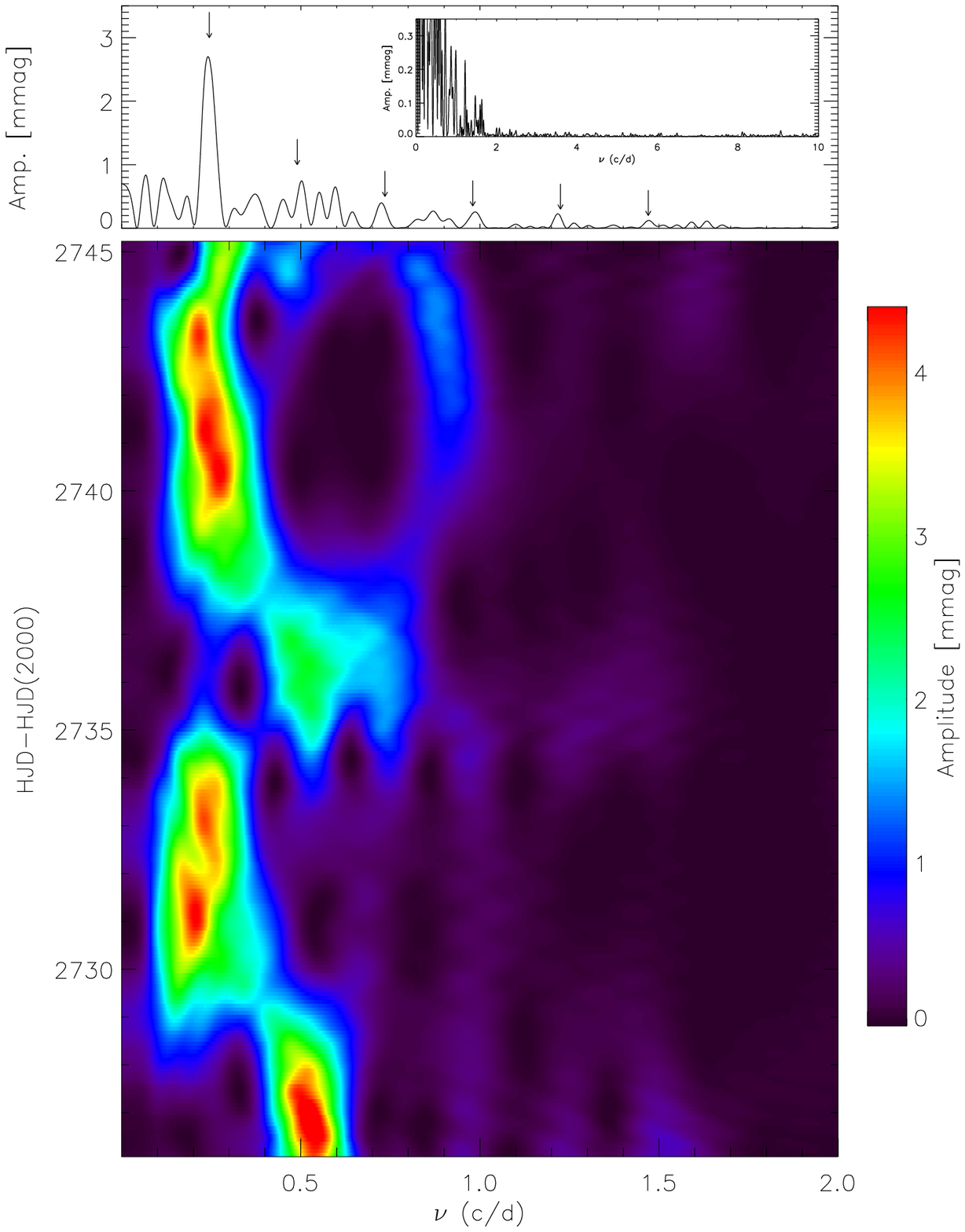}
 \caption{Upper panel: Fourier amplitude spectrum of the binned {\it MOST} light curve in Figure~\ref{fig1}, with insert showing expanded scales. The highest peak refers to the adopted fundamental frequency corresponding to a period of P~=~4.08~days.  Harmonics are indicated at frequencies at multiples of 2 through 6 times fundamental (marked with downward arrows). Lower panel: Time-frequency Fourier plot with 8-day running windows in time.}\label{fig3}
\end{figure}

We start by examining the Fourier spectrum for the {\it MOST} photometry (Figure~\ref{fig3}). Here the largest amplitude occurs at a frequency of $\nu_0 =$ 0.245 $\pm$ 0.033 cycles day$^{-1}$ - with error equal to $1/T_{ph}$, where $T_{ph}$ is the total duration of the light curve - (P = 4.08 $\pm$ 0.55 days), with harmonics of this frequency at, or close to, $n \times \nu_0$, with $n\approx$ 2, 3, 4, 5 and 6. A few other peaks do occur at low ($<$20\% of the dominant peak) relative amplitudes, notably at periods longer than $\sim$a day, but with no obvious order or interrelationship. None are due to sidelobes of the primary maximum, due to the uniform, dense data sampling. No significant peaks beyond the above harmonics were detected with periods shorter than $\sim$a day. We assume that this implies a single dominating (non-sinusoidal) periodicity with P = 4.08~days, plus mainly stochastic variability on timescales longer than $\sim$a day \citep[possibly related to wind clump activity, which survives for at least $\sim$8 hours in typical WR winds: ][]{Le99}.  Examination of the light curve in Figure~\ref{fig1} shows relatively sharp cusps occurring on this time scale (i.e. P = 4.08 days, starting at $t_0$ = HJD 2~454~270.27) in all seven cases but one (the third case), although even then there is a rise in brightness, though not cusp-like. The cusps have typical widths at the base of $\sim$a day.  Such cusps are clearly non sinusoidal, requiring a series of successive harmonics to reproduce. Figure~\ref{fig3} also shows a time-frequency plot with straddled 8-day (double the best period) sampling. This plot reveals that the 4.08-day period stands out at all times except near the middle and at both ends (where edge effects come into play) of the photometric time series. The former is compatible with the apparently suppressed third cusp, as noted above.

\begin{figure}[ht]
 \plotone{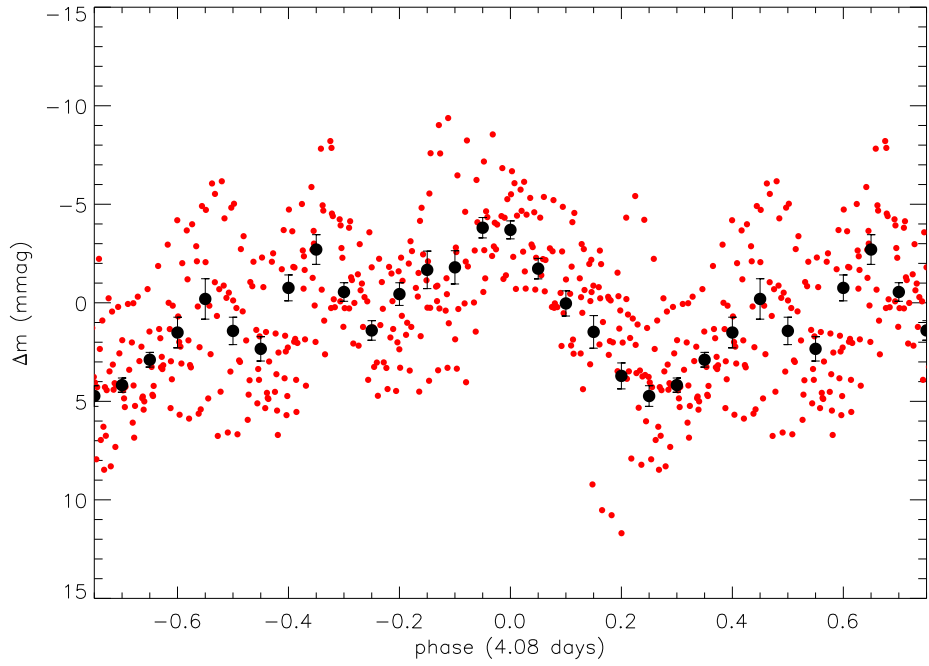}
 \caption{Phased light curve from Figure~\ref{fig1}, using P = 4.08 days and $t_0 =$ HJD~2454270.27. The large points show means in phase bins of 0.05, with 2-$\sigma$ error bars based on the standard error of the mean in each bin.}\label{fig4}
\end{figure}

Figure~\ref{fig4} shows the phase-binned light curve of the orbital-mean {\it MOST} data from Figure~\ref{fig1}, based on the period of 4.08~days and zero point $t_0$ (nearly coinciding with the first observed cusp in Figure~\ref{fig1}). This light curve is quite noisy compared to the appearance of the non phase-binned light curve in Figure~\ref{fig1}.  Further binning to 0.05 in phase reveals a dominating peak at phase zero, with a second lower peak at phase $\sim$0.65. Both peaks can be traced back in Figure~\ref{fig1} often to sharper, clearer cusps at similar repeated intervals. These cusps become somewhat smeared, although remain still visible in Figure~\ref{fig4}, probably due to dispersion in shape and timing around some average, combined with stochastic wind variations.

\subsection{Spectral Line Variations}

\subsubsection{Short-term}

\begin{figure}[ht]
 \plottwo{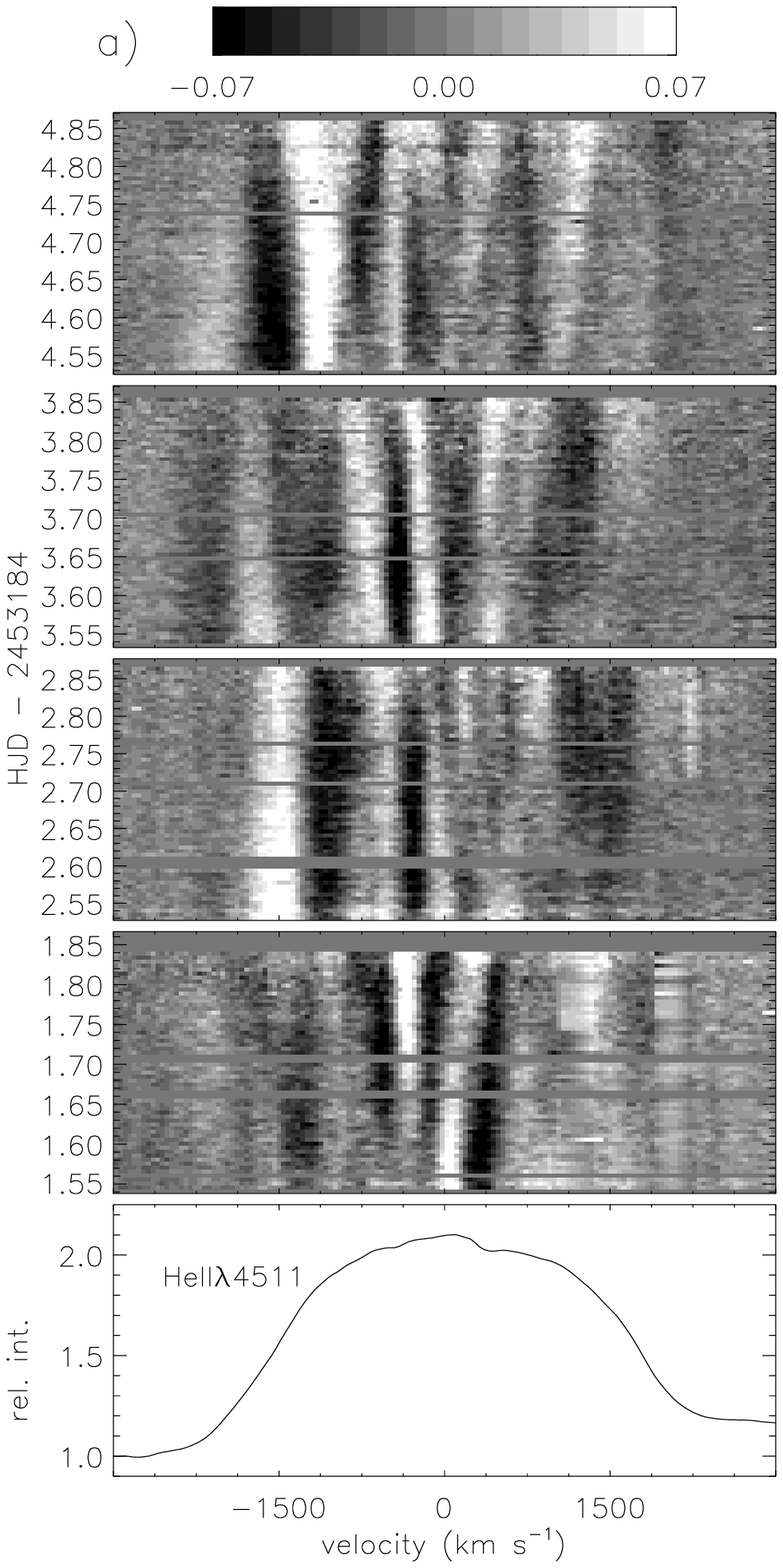}{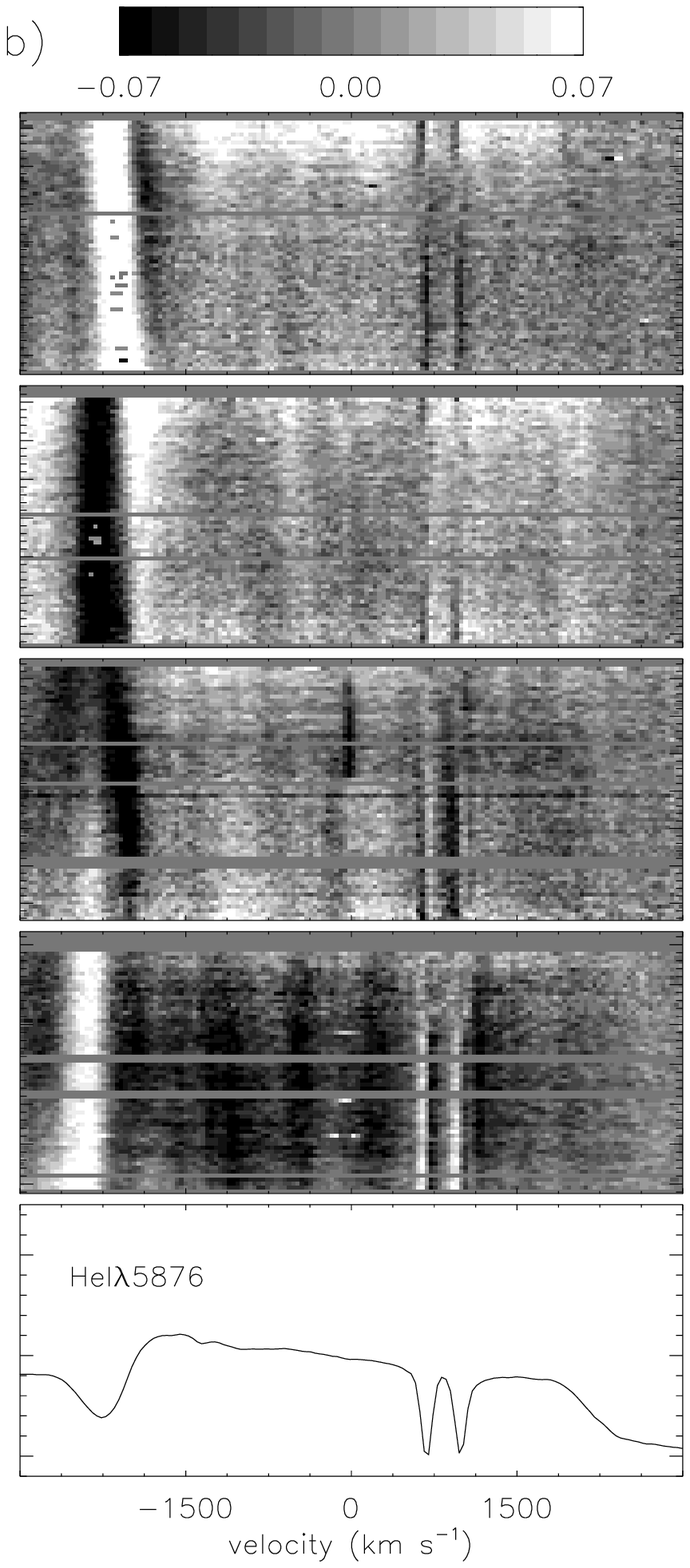}
 \caption{Upper 4 panels: Grey-scale plots of the difference from the mean of the CTIO spectra over four successive nights (shown in the bottom panel). (a) He{\sc ii}$\lambda$5411, (b) CIV{\sc iv}$\lambda$5802/12 + He{\sc i}$\lambda$5876.}
  \label{fig5}
\end{figure}

Only one spectroscopic run (CTIO 2004) is of sufficient quality and length to allow a search for variability on timescales less than a day. Figure~\ref{fig5} shows greyscale time-plots of the four successive CTIO nights for differences from the mean of each spectrum, for the principal WR spectral emission lines of He{\sc ii}$\lambda$5411 and C{\sc iv}$\lambda\lambda$5802/12 + He{\sc i}$\lambda$5876. From these plots, one sees evidence for the usual stochastic, emission subpeaks, propagating outwards from line center, as seen in all previously well-studied WR spectra \citep{Le99}. Such perturbations are believed to arise in turbulent wind clumps, with no apparent periodicity on timescales of a night ($\sim$0.3 day) or shorter.

The simplest, most effective way to examine spectral line variations is via time dependence of line moments from order zero through four.  The zeroth moment gives the line strength or equivalent width (EW) when normalized to the continuum.  The first moment yields the global line position or radial velocity (RV). The second moment can be used to calculate the line width \`a la gaussian.  Line asymmetries can best be estimated by (normalized) third moments, known as skewness.  Finally, normalized fourth moments yield information on how the profile differs in global shape from a gaussian.

\begin{figure}[ht]
 \plotone{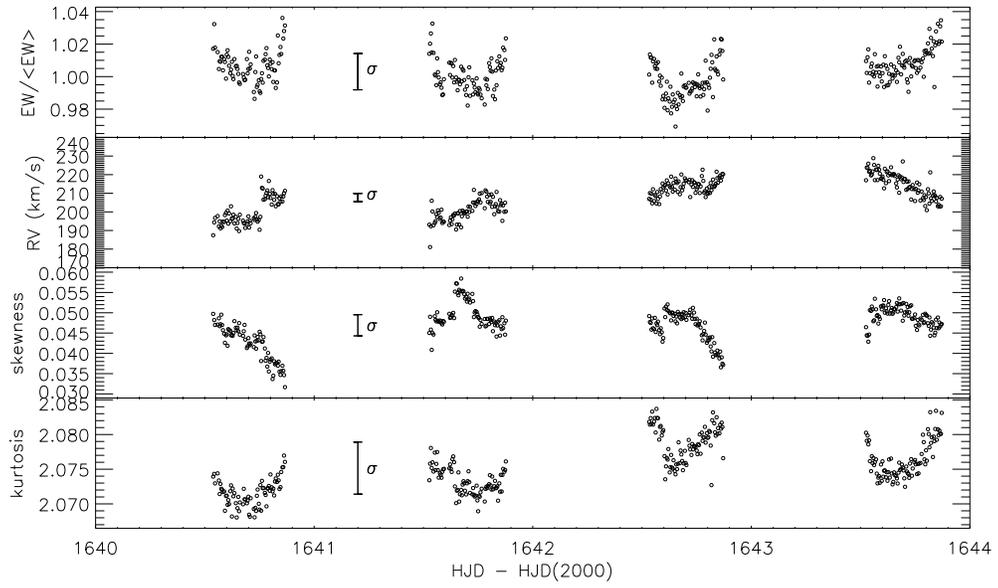}
 \caption{Normalized equivalent width (EW), line radial velocity (RV), skewness and kurtosis versus time for the CTIO data of He{\sc ii}$\lambda$5411.}
  \label{fig6}
\end{figure}

Figure~\ref{fig6} shows a time plot of normalized EW, RV, skewness and kurtosis of the CTIO observations for He{\sc ii}$\lambda$5411. The EWs were calculated by integrating the function $(1-F_\lambda)$, where $F_\lambda$ is the rectified line flux between $\Delta \lambda$=5348--5460\,\AA. In order to highlight the variability, we prefer to divide the EWs by the mean value of all EWs. The radial velocities (RVs) were obtained by cross correlating
individual spectra in the wavelength interval $\Delta \lambda$=5140-5980\,\AA\, with the mean spectra. (Line bisectors were also explored to obtain RVs, but they revealed larger scatter yet similar trends as the cross-correlation technique.) The other moments of the He{\sc ii}$\lambda$5411 line are calculated in the wavelength interval $\Delta\lambda$=5348-5460\,\AA. The $n^{th}$ central moment is defined as follows:
\begin{eqnarray}
\mu_n=\Sigma_j(\lambda_j-\bar{\lambda})^n I_j/\Sigma_jI_j
\end{eqnarray}
where
\begin{eqnarray}
\bar{\lambda}=\Sigma_j\lambda_jI_j/\Sigma_jI_j,
\end{eqnarray}
with $I_j$ being the intensity of the line and $\lambda_j$ the wavelength. The skewness is $\mu_3/\mu_2^{3/2}$ and the kurtosis is $\mu_4/\mu_2^{2}$. Variability is seen in all of these quantities, although with no significant periodicities. Any longer-term periodic variations of these data will be examined together with the remaining spectra in the next subsection.

\subsubsection{Long-term}

\begin{figure}[ht]
 \plotone{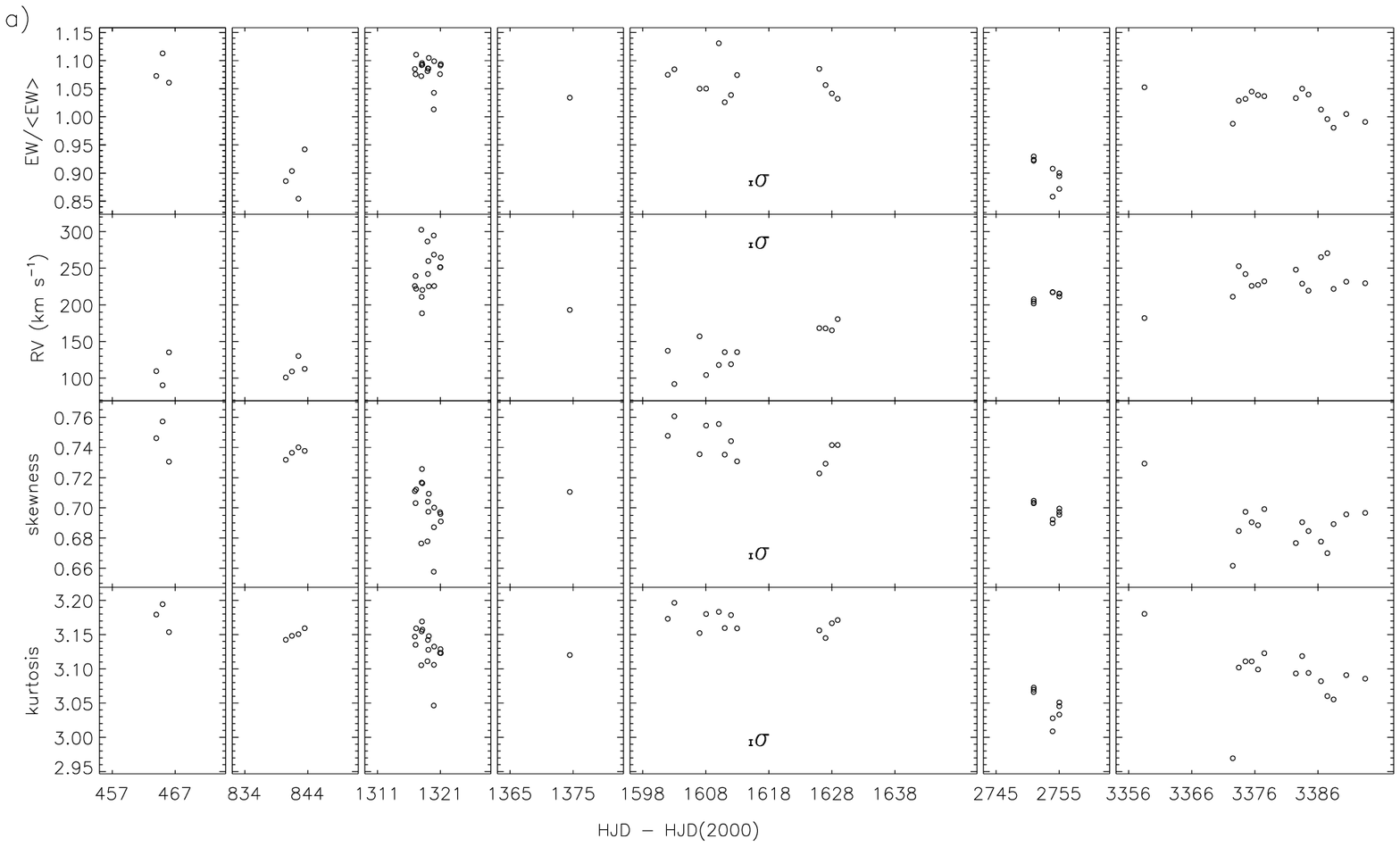}
 \plotone{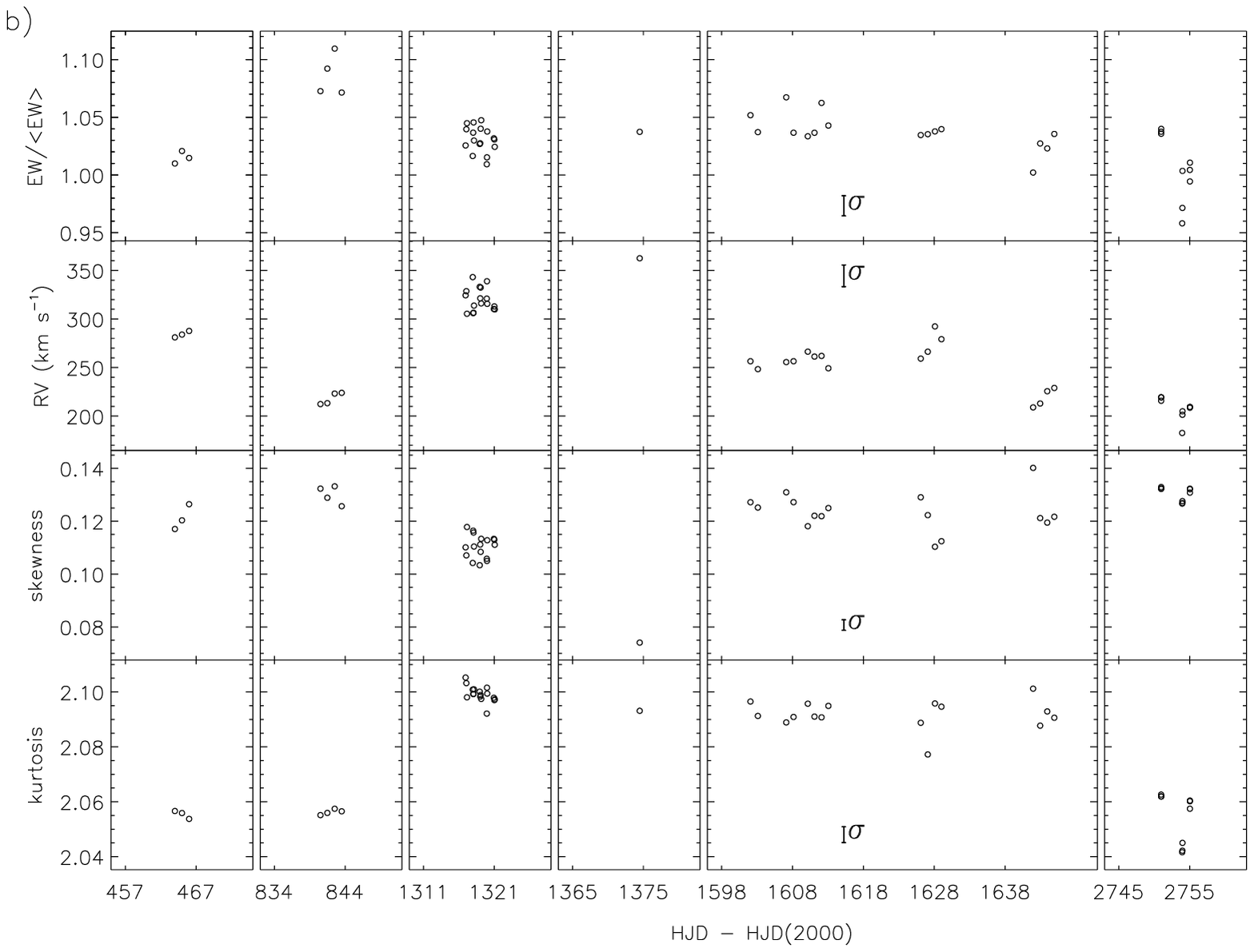}
 \caption{As in Figure~\ref{fig6} for all the spectroscopic runs for (a) He{\sc ii}$\lambda$4686 and (b) He{\sc ii}$\lambda$5411, with EW relative to the global mean (excluding the OMM data, due to detector zero-point intensity problems). The CTIO data for He{\sc ii}$\lambda$5411 are shown in nightly means.}\label{fig7}
\end{figure}

Figure~\ref{fig7} shows time plots of EW, RV, skewness and kurtosis (measured as above) for the strongest, least blended lines: (a) over the full 6 years for He{\sc ii}$\lambda$4686 and (b) $T_{sp} =$ 8 years for He{\sc ii}$\lambda$5411. In the case of the latter from CTIO, we have taken nighly averages, so that the numerous original CTIO data will not dominate the period analysis. The moderately strong lines of C{\sc iv}$\lambda$5802/12 and He{\sc i}$\lambda$5876 are blended and seriously affected by telluric lines, so they were not analyzed.

While these four parameters all show variations for both lines, those of largest amplitude are in RV and skewness for He{\sc ii}$\lambda$4686. These two quantities appear to vary in an anti-correlated way (correlation amplitude, r = -0.89), as expected if the line varies in shape; e.g. an excess on the red side of the line will cause a small positive RV shift and a negative change in skewness. A similar trend (r = -0.86), is also measured for these two parameters in the He{\sc ii}$\lambda$5411 line, despite the apparently lower amplitude of the variations. Presumably, these are variations in profile shape and do not reflect true RV motions of the star. Furthermore, a search using phase-dispersion modulation \citep[PDM,][]{St78} and Fourier methods \citep{Be04} was carried out for periodicities in the RVs in the range expected from the {\it MOST} variations This was done in the frequency range $\nu$ = 0.2037 - 0.2885 cycles day$^{-1}$, covering the whole photometric Fourier peak centered at P = 4.08 days, and with frequency step size $2\times10^{-5}$ cycles day$^{-1}$, i.e. $\sim (10\times2T_{sp})^{-1}$. The search failed to reveal any significant periodicity. Alternatively, a phase RV plot of each line phased with the best period in this interval produces a scatter diagram with force-fitted sine-wave RV amplitude of K $\sim$14, 10 and 28~km~s$^{-1}$ for all RV data (nightly means for CTIO, with typical errors of $\sim$10 km s$^{-1}$) for He{\sc ii}$\lambda$5411, the 335 CTIO data alone for He{\sc ii}$\lambda$5411, and all RV data for He{\sc ii}$\lambda$4686, respectively. The lower the scatter, the lower the K value. Combined with the above information on anti-correlation of RV and skewness, we deduce that if WR\,110 is a binary system, it is not evident from the data presented here. On the other hand, the lack of a clear indication of a 4.08 day period in any of the spectroscopic data does not necessarily mean that such a period is not present. We suspect that it will take a much larger, higher-quality spectroscopic sample to reach any firm conclusions on any periodicity in spectral behavior of WR110, as was the case with WR\,134 \citep{Mo99}.

\section{CIR model}

Plausible scenarios to account for the 4.08 day periodicity include binarity,  pulsations and CIRs. Binarity is unlikely, due mainly to a lack of coherent RV variations that phase with the primary photometric period. Pulsations are unlikely due to the highly non-sinusoidal form of the variations at P = 4.08 days. Hence, the most likely  scenario to explain the low-amplitude ($\sim$1 \%) 4.08 day periodicity and cusp-like behavior in the {\it MOST} light curve, along with mostly unrelated spectral variations, appears to be large-scale, emitting over-density structures rotating with the wind. (One also has the impression of cusp-like minima in the light curve; however, these could be a result of the small space between two emitting cusps occurring per rotation - see below.) This would be similar to what is seen in a few other WR stars (WR6, WR134, WR1 - see below). Corotating interaction regions (CIRs; or their source at the base of the wind) as proposed for O star winds by \citet{Cr96} provide an attractive scenario to explain our observations. The Cranmer \& Owocki model is designed for O stars, explicitly assuming CAK \citep{Ca75} parameters, which may not apply directly to H-free WR stars.  Nevertheless, the idea of a CIR driver at the base of the wind in WR stars is a qualitatively plausible scenario, analogous to O-star winds. We now apply this idea to WR\,110 in the following way.

First, in the context of the Cranmer \& Owocki model, we assume that the CIRs are continuum-emitting thermal sources, heated by the associated shock action created by a hot spot on the underlying rotating star interacting with the ambient wind. Most of the CIR emission must arise close to the star, where the ambient wind density is highest, as inspired by the simulations of \citet{Cr96} - see a simulation in their Fig. 5. Alternatively, the emission could arise in a hot spot at or close to the stellar surface, that produces the CIR. To simplify our calculations, we assume a point source located at some radius $R_s =\gamma R_*$ from the center of the star, with $\gamma > 1$ and $R_*$ being the stellar core radius. For simplicity \citep[and following][]{Cr96} we assume the CIR to initiate at the stellar rotation equator. 

In this scenario, the changes in the light-curve are caused by an assumed CIR-associated point source attenuated by the wind and seen at different angles with the line of sight as it rotates on the near side of the WR star. In this study, we use a simple WR wind model derived by \citet{La96} in the context of wind eclipses for WR + O systems. Hence, the change in magnitudes is defined as~:
\begin{eqnarray}\label{dm}
\Delta m\,=\,constant\,-\,2.5 \log_{10}\left(I_{WR}+I_s \rm{e}^{-\tau}\right),
\end{eqnarray}
with arbitrary constant and where $I_{WR}$ is the intensity of the WR
star and $I_s$ is the intensity of the hot spot. Here the total opacity
$\tau$ between the source and the observer, passing through the WR
wind is~:
\begin{eqnarray}\label{tau}
\tau=k\int_{z_0}^\infty{d\left(z/R_s\right)\left[(r/R_s)^2\left(1-R_*/r\right)^\beta\right]^{-1}},
\end{eqnarray}
where $(r/R_s)^2 = (\cos{i} \cos{2 \pi \phi})^2 + \sin^2{2 \pi \phi} + (z/R_s)^2$,  $z_0=-(R_s \sin{i}) \cos{2\pi\phi}$, $i$ is the inclination of the rotation axis (assumed to coincide with the CIR rotation axis) relative to the observer, and $\phi$ is the rotation phase. We allow for a WR wind with a $\beta$ law \citep[like in ][]{La96}, in which the actual $\beta$ value will be fitted. Since the CIR source is much closer to the inner WR wind than the O star is in the (non photospheric eclipsing) binary case, we expect that $\beta = 0$ will not be a good choice, as it was in the case of WR + O binaries \citep{La96}. Following \citet{Sk02}, we adopt a radio-based thermal mass-loss rate for the WR star of $\dot{M} =$ 1.6 10$^{-5}$ M$_{\odot}$\,yr$^{-1}$, after dividing by a factor three to allow for wind clumping, a wind terminal speed of $v_{\infty} =$2100~km~s$^{-1}$, and a WR core radius $R_* =$ 4 $R_{\odot}$. We find an opacity constant as in \citet{La96} $k = \alpha \dot{M} \sigma_{e}/[4 \pi m_p v_\infty R_s] = 0.28/\gamma$, where $\alpha =$ 0.5 is the number of free electrons per He nucleus in a highly ionized wind. In our case, we are primarily interested in the phases centered at phase 0.50 (i.e. when the source is between the star and the observer) since the CIR point source will not be normally visible as it rotates on the back side of the WR star.

We calculated Equation~\ref{dm} numerically, centered at phase 0.5 for half a rotation period, as a function of the parameters $\beta$, $I_s/I_{WR}$, $i$, and $\gamma$. We consider two cases: (a) an optically thin point source, whose emission is isotropic in direction, and (b) an optically thick source whose emission is maximum when seen perpendicularly to the stellar surface, dropping to zero when seen parallel to the surface (i.e. like pancakes on or near the stellar surface). In case (a) light-curve excesses originate from the hot spot and in case (b), we have to multiply the net spot intensity ($I_s$) from Equation~\ref{dm} by the projection factor $(\sin i) \cos[2\pi(\phi - 0.5)]$.

\begin{figure}[ht]
 \plotone{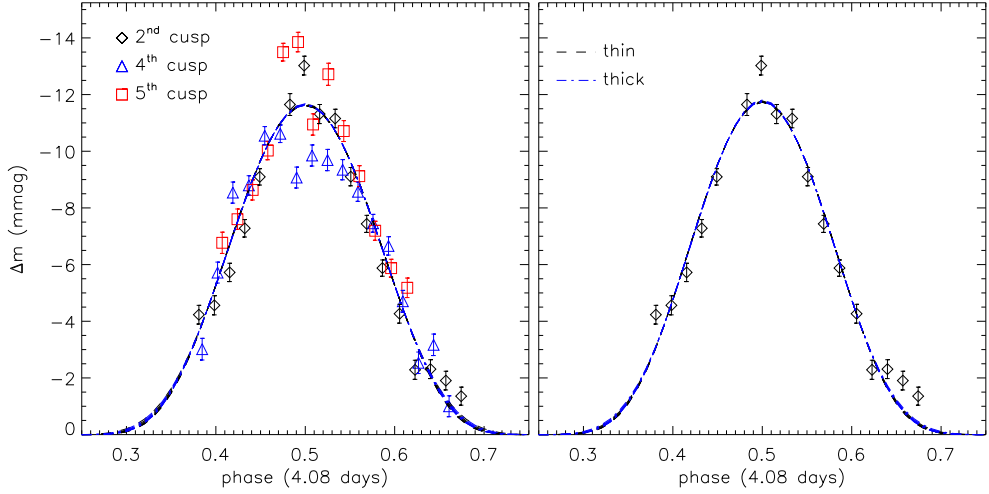}
 \caption{{\it Left}:The data points refer to the second (black diamonds), the forth (blue triangles) and the fifth (red squares) cusps indicated by vertical lines in Figure~\ref{fig1}. We assume that these cusps begin and end when the light curve reaches a minimum, i.e. when there is no enhancement due to the passage of a CIR. To avoid contamination from previous and subsequent features, three cusps are selected in the time intervals HJD~2~454~273.725 -- 275.000, HJD~2~454~282.240 --  283.450 and HJD~2~454~285.770 -- 286.720, respectively. To facilitate the fit, the data points are shifted vertically by an arbitrary constant, so the minimum is fixed at $\Delta m$=0. The curves indicate the best fits at different inclination angles (see Table~\ref{tab2}) for an optically thin point source, using black dashed lines, and an optically thick source, using blue dash-dotted lines (see text). Note that these curves are superposed and hard to distinguish {\it Right}: Same as the left panel, but only using the second cusp.}\label{fig8}
\end{figure}

Taking the second, the fourth and the fifth cusps in Figure~\ref{fig1} as the cleanest, most representative form of the hotspot light curve, we fit the above parameters using the IDL routine {\it mpfit.pro} \citep{Ma09}. The fit does not converge when the inclination $i$ is set as a free parameter. Therefore, we fix it to different angles in the range $i = 90^\circ-40^\circ$ with increments of $10^\circ$ before fitting all the other parameters. A summary of fitted parameters using different inclination angles is given in Table~\ref{tab2}. Note that we do not include our results for an inclination $i$ lower than $40^\circ$, since the fit does not converge when using such small angles. The values of $I_s/I_{WR}$ tend to be slightly smaller in case (b) than in case (a), due to the projection factor in case (b) that requires less extended wind opacity to bring about the observed cuspy shape. Indeed, it is the varying opacity through the WR wind as the WR star rotates, which leads to the cuspy shape; the closer the source is to the stellar surface (i.e. $\gamma$ closer to unity), the smaller the $\beta$ value (and less extended the wind) necessary to give the cuspy shape. The best solutions for the thin and the thick cases at different inclination angles  are plotted in the left panel of Figure~\ref{fig8}. We also repeated the fits for the thin and thick cases at the different inclination angles using only the second cusp, in order to verify how our results would vary depending on this choice. The best solutions are plotted in the right panel of Figure~\ref{fig8} and also summarized in Table~\ref{tab2}. The difference between the results using one or three cusps may come from an epoch-dependency of the variations, smaller cusps present in only some of the considered cusps and/or a change in the shape and amplitudes between the different cusps due to superposed stochastic variability.

All the fits give a $\chi^2_{red}$ close to unity. It is hence, at first, not possible to exclude any result only based on the quality of the fit. As seen in Figure~\ref{fig8}, changing the inclination does not change the shape of the curve as much as the amplitude and requires higher values of $\beta$, $\gamma$ and $I_s/I_{WR}$ (this may explain why the fit does not converge when the inclination $i$ is set as a free parameter). However, in many cases $I_s/I_{WR} > 1$, which means that the intensity of the spot is greater than the intensity of the whole star. Since such a solution is not likely, we may exclude the optically thin case, and set $i\ga70^\circ$. The remaining solutions imply $\beta\approx11.5-12.5$. These value would give $\beta R_\ast \approx50 R_\odot$, which is comparable to the value obtained by \citet{Le99} for stars of similar spectral type.

Although the overall fit is quite good, the model does not reproduce well the peaked central cusps. Tweaking the model to do this may require assumptions that go beyond the scope of this study. For example, it might be possible that the expected density-deficit channel parallel to the CIR \citet{Cr96} could affect the light-curve peaks. 

\section{Discussion and Conclusions}

Assuming that CIRs do in fact account for the variability observed in the {\it MOST} light curve of WR\,110, we can now provide a plausible explanation for the hard X-rays coming from WR\,110. According to the models of CIR \citep{Cr96}, shock velocities in the radial direction can reach over 10$^3$ km s$^{-1}$ at several stellar radii (beyond where the optical hot spot is seen), although highly variable from one model to another. This is several times larger than the velocity dispersion \citep[$\sim\, 10^2$~km~s$^{-1}$:][]{Le99} in clumps that is likely associated with the low-energy X-rays (kT $\approx$ 0.5 keV). Scaling up proportional to the square of the velocity, one can easily account for the observed hard X-rays using CIRs without accretion onto a compact or low-mass binary companion. Note that the wind terminal velocity of WR\,110, although high \citep[2100~km~s$^{-1}$ from][or 2300~km~s$^{-1}$ from \citealp{Ha06}]{Sk02}, is not the source per se of the X-rays; it is the velocity {\it dispersion} or the radial CIR velocity component which counts for producing the low- versus high-energy X-ray shocks, respectively. 

Our modeling of the photometric light curve leads us to conclude that WR\,110 joins the ranks of three other WR stars, all WN, that exhibit CIRs with known rotation periods of the underlying star: WR\,6, WN4, P = 3.76 days; WR\,134, WN6, P = 2.3 days; WR\,1, WN4, P = 16.9 days \citep[][respectively]{Mo97,Mo99,Ch10}. While the periodicity in WR\,6 was easy to find because of the high photometric amplitude ($\sim$ 0.1 mag), WR\,134 was more difficult with $\sim$ 0.03 mag plus significant noise from other sources, and despite its large amplitude $\sim$ 0.1 mag, WR\,1 has a long period and complex light curve.  WR\,110 has an even lower amplitude, with significant noise, making the availability of {\it MOST} for intense, long-term monitoring indispensable for the discovery of its CIRs. In all 4 cases, the cuspy nature of the often multiple light-curve bumps, is evident and may thus be due to the same explanation: continuum-emitting CIRs that emit mainly just outside the inner opaque WR wind, where the intensity falls off rapidly at larger radii. In most cases, there are 2 cusps, implying 2 CIRs. This is the usual number found in O-star winds \citep{Ka99}.

None of these rotation rates appear to be rapid enough to be directly associated with Gamma-ray Bursts, assuming rotational coupling between the surface and the interior \citep{St09}.

Despite its very small amplitude and complex pattern, WR\,110 has possibly revealed its CIR rotation properties due to the power of the {\it MOST} satellite, i.e. its ability to observe continuously for a month at high precision. This may open up the door to revealing other subtle cases, giving us a better idea of the properties of WR stars and their strong winds. It may even have given us a lead to probing the mysterious origin of the CIR phenomenon rooted at the stellar surface, e.g. as a phenomenon related to magnetism or pulsations.

\acknowledgments

ANC gratefully acknowledges support from the Chilean {\sl Centro de Astrof\'\i sica} FONDAP No. 15010003 and the Chilean Centro de Excelencia en Astrof\'\i sica y Tecnolog\'\i as Afines (CATA). DBG, JMM, AFJM, SMR and NSL are supported by NSERC (Canada), with additional support to AFJM and NSL from FQRNT (Qu\'ebec). RK and WWW are supported by the Austrian Space Agency and the Austrian Science Fund. KZ acknowledges support from the Austrian Fonds sur F\"orderung der wissenschaftlichen Forschung (FWF) and is recipient of an APART fellowship of the Austrian Academy of Sciences at the Institute of Astronomy of the University of Vienna. We thank the anonymous referee for useful suggestions to improve the clarity of this work.

\clearpage

\include{tab1}
\include{tab2}

\end{document}

%% file: tab1.tex
\begin{table}
\begin{center}
\caption{Journal of spectrosocpic observations.\label{tab1}}
\begin{tabular}{llccc}
\tableline\tableline
Telescope & Dates & No. spectra & Range (\AA) & \AA/pi \\
\tableline
Leoncito 2.2m & 2001 Apr 8 -- 10      & 3 & 3900-5500 & 1.63\\
              & 2002 Apr 20 -- 23 & 4 &     ''      &  ''   \\
              & 2003 Oct 5     & 1 &     ''      &   ''  \\
              & 2004 May 20 -- 31 & 8 &     ''      &  ''   \\
              & 2004 Jun 13 -- 16 & 4 &      ''     &  ''   \\
CFHT 3.6m     & 2003 Aug 9 -- 13  & 17 & 4300-6000 & 0.51\\
CTIO 1.5m     & 2004 Jun 29 -- Jul 2 & 335 & 5300-6200 & 0.77\\
OMM 1.6m      & 2007 Jul 13 -- 17   & 5 & 4400-6100 & 0.66\\
Leoncito 0.6m & 2009 Mar 12 -- Apr 16 & 15 & 3700-5300 & 0.97\\ 
\tableline
\end{tabular}
\end{center}
\end{table}

%% file: tab2.tex
\begin{table}
  \begin{center}
    \caption{CIR model fit parameters. \label{tab2}}
      \begin{tabular}{ccccccc}
        \tableline\tableline
        Optically & nb. cusps & inc ($^\circ$) &$\beta$ & $\gamma$ & $I_s/I_{WR}$ & $\chi^2_{red}$\\
        \tableline
        Thin  & 3 & 90 & 13.36 & 2.825 & 1.351 & 1.489 \\
              &   & 80 & 13.39 & 2.832 & 1.393 & 1.484 \\
              &   & 70 & 13.49 & 2.853 & 1.536 & 1.471 \\
              &   & 60 & 13.67 & 2.889 & 1.842 & 1.447 \\
              &   & 50 & 13.98 & 2.947 & 2.497 & 1.412 \\
              &   & 40 & 14.48 & 3.032 & 4.145 & 1.363 \\
        \tableline
        Thick & 3 & 90 & 11.46 & 2.652 & 0.405 & 1.442 \\
              &   & 80 & 11.47 & 2.656 & 0.421 & 1.440 \\
              &   & 70 & 11.59 & 2.681 & 0.476 & 1.430 \\
              &   & 60 & 11.76 & 2.716 & 0.594 & 1.412 \\
              &   & 50 & 12.08 & 2.778 & 0.847 & 1.386 \\
              &   & 40 & 12.56 & 2.863 & 1.488 & 1.350 \\
        \tableline
        Thin  & 1 & 90 & 14.19 & 2.895 & 2.595 & 1.198 \\
              &   & 80 & 14.20 & 2.899 & 2.685 & 1.194 \\
              &   & 70 & 14.30 & 2.920 & 2.993 & 1.180 \\
              &   & 60 & 14.51 & 2.960 & 3.666 & 1.155 \\
              &   & 50 & 14.81 & 3.017 & 5.151 & 1.119 \\
              &   & 40 & 15.32 & 3.102 & 9.095 & 1.068 \\
        \tableline
        Thick & 1 & 90 & 12.48 & 2.744 & 0.779 & 1.154 \\
              &   & 80 & 12.51 & 2.750 & 0.813 & 1.150 \\
              &   & 70 & 12.62 & 2.773 & 0.929 & 1.140 \\
              &   & 60 & 12.79 & 2.808 & 1.187 & 1.121 \\
              &   & 50 & 13.12 & 2.870 & 1.758 & 1.094 \\
              &   & 40 & 13.62 & 2.956 & 3.294 & 1.055 \\
        \tableline\tableline
      \end{tabular}
  \end{center}
\end{table}